\documentclass[preprint,preprintnumbers,amsmath,amssymb]{revtex4}
\usepackage{amsthm}
\usepackage[colorlinks=true,urlcolor=blue,citecolor=blue,linkcolor=blue]{hyperref}
\usepackage[shortlabels]{enumitem}
\usepackage{amsfonts}
\usepackage[caption = false]{subfig}
\usepackage{multirow}
\usepackage{longtable}
\usepackage{float}
\usepackage{bbold}
\usepackage{stmaryrd}
\usepackage{color}
\usepackage{hyperref}
\usepackage{verbatim}
\usepackage{cases}
\usepackage{amsfonts}
\usepackage{amssymb}
\usepackage[]{mathrsfs}
\usepackage{xcolor}           
\usepackage{epsfig}
\usepackage{setspace}
\usepackage{enumerate}
\usepackage{dcolumn}
\usepackage{bm}
\usepackage{enumitem}
\usepackage{multirow, array} 
\newcommand{\pder}[2]{\frac{\partial #1}{\partial #2}}
\usepackage{dcolumn}
\usepackage{bm}
\usepackage{color}
\usepackage{amssymb}
\usepackage{lipsum}
\usepackage{amsmath}
\usepackage{array}
\usepackage{multirow}
\usepackage{booktabs}

\usepackage[english]{babel}

\begin{document}


\title{ 3-body harmonic molecule}

\author{H. Olivares-Pil\'on}%
\email{horop@xanum.uam.mx}
\affiliation{Departamento de F\'{i}sica, Universidad Aut\'onoma Metropolitana Unidad Iztapalapa, San Rafael Atlixco 186, 09340 Cd. Mx., M\'exico}

\author{A. M. Escobar-Ruiz}
\email{admau@xanum.uam.mx}
\affiliation{Departamento de F\'{i}sica, Universidad Aut\'onoma Metropolitana Unidad Iztapalapa, San Rafael Atlixco 186, 09340 Cd. Mx., M\'exico}

\author{F. Montoya}%
\affiliation{Departamento de F\'{i}sica, Universidad Aut\'onoma Metropolitana Unidad Iztapalapa, San Rafael Atlixco 186, 09340 Cd. Mx., M\'exico}



\begin{abstract}
In this study, the quantum 3-body harmonic system with finite rest length $R$ and zero total angular momentum $L=0$ is explored. It governs the near-equilibrium $S$-states eigenfunctions $\psi(r_{12},r_{13},r_{23})$ of three identical point particles interacting by means of any pairwise confining potential $V(r_{12},r_{13},r_{23})$ that entirely depends on the relative distances $r_{ij}=|{\mathbf r}_i-{\mathbf r}_j|$ between particles. At $R=0$, the system admits a complete separation of variables in Jacobi-coordinates, it is (maximally) superintegrable and exactly-solvable. The whole spectra of excited states is degenerate, and to analyze it a detailed comparison between two relevant Lie-algebraic representations of the corresponding reduced Hamiltonian is carried out. At $R>0$, the problem is not even integrable nor exactly-solvable and the degeneration is partially removed. In this case, no exact solutions of the Schr\"odinger equation have been found so far whilst its classical counterpart turns out to be a chaotic system. For $R>0$, accurate values for the total energy $E$ of the lowest quantum states are obtained using the Lagrange-mesh method. Concrete explicit results with not less than eleven significant digits for the states $N=0,1,2,3$ are presented in the range $0\leq R \leq 4.0$~a.u. . In particular, it is shown that (I) the energy curve $E=E(R)$ develops a global minimum as a function of the rest length $R$, and it tends asymptotically to a finite value at large $R$, and (II) the degenerate states split into sub-levels. For the ground state, perturbative (small-$R$) and two-parametric variational results (arbitrary $R$) are displayed as well. An extension of the model with applications in  molecular physics is briefly discussed. 
\end{abstract}

\maketitle



\section{Introduction} 

The quantum system of $n$ ($n>2$) point particles in $\mathbb{R}^3$ with arbitrary masses connected through springs can be considered as a natural $n$-body generalization of the celebrated two-body harmonic oscillator ($n=2$), the latter being a system of tantamount relevance in theoretical physics \cite{moshinsky1996harmonic} and intimately close to with the method of second quantization introduced by Dirac in the late 1920s. The corresponding potential $V(r_{ij})$ in classical and quantum mechanics, a linear combination of the squares of the mutual relative distances $r_{ij}=|{\mathbf r}_i-{\mathbf r}_j|$, is superintegrable and exactly-solvable. However, from a physical point of view a more adequate model is obtained by the replacement $r_{ij}^2 \rightarrow {(r_{ij}-R_{ij})}^2$ where the constants $R_{ij}>0$ play the role of rest lengths. This model can be called generalized $n$-body harmonic system (GNBHS). Despite of its apparent modesty, in the simplest three-body case $n=3$ not a singly exact solution of the Schr\"odinger equation is known so far. 

In classical mechanics, the three-body generalized harmonic system $n=3$ possesses a complex rich dynamics. For the case of identical particles, with common rest length $R_{ij}=R>0$, a chaotic behaviour as a function of the energy and the system parameters occurs \cite{PhysRevLett.122.024102}-\cite{PhysRevE.101.032211}.  At fixed energy $E$, and restricted to the invariant manifold of zero total angular momentum, its dynamics transits between two regular regimes passing through a chaotic one as the parameter $R$ grows. Recently, an experimental physical realization of this system was built by means of analog electrical components \cite{Classicalcase}. Therefore, the GNBHS represents a suitable candidate to test theoretical and numerical tools to analyze the nature of the classical-quantum relation in $n-$body chaotic systems \cite{Gutzwiller1971PeriodicOA}.

In quantum mechanics, the solutions of the GNBHS are of 
great theoretical importance. Moreover, in practice they could be used 
as a basis for many-body calculations in molecular,  nuclear 
and elementary particle physics just to mention few examples~\cite{JMR:1992}. 

In the present work, for the $3$-body generalized harmonic system we now ask how the energies and eigenfunctions of the quantum system behave as the rest length $R$ increases. We restrict ourselves to the symmetric case of 3 identical particles $m_1=m_2=m_3=m$ with zero total angular momentum $L=0$ ($S-$states). The unfolding of the degenerate states due to a non zero value of $R$ is of particular interest. It is worth mentioning that different aspects of the three-body system in $\mathbb{R}^d$ $(d>1)$ with arbitrary masses ($m_1,m_2,m_3$) have been presented in \cite{Turbiner_2017,doi:10.1063/1.4994397} (see also \cite{Gu}). For instance, a reduced Hamiltonian for the $S$-states which solely depends on the 3 coordinates $r_{ij}$ was established. Also, at $R=0$ a complete analysis of the one-dimensional case $d=1$ (three masses on a line) can be found in \cite{FF}. Here, for 3 equal masses in $\mathbb{R}^3$ and $R\neq 0$, using the Lagrange-Mesh Method the solutions of the lowest $S-$states eigenvalues of the corresponding Schr\"odinger equation are computed with high accuracy. Specifically, the energies are displayed with not less than eleven significant figures. 

The structure of the paper is as follows. In Section II we define the generalized $3$-body harmonic system  and the concrete setting of the problem is explained. Especially, the relevant reduced Hamiltonian governing the $S$-states is described. At zero rest length $R=0$, the system becomes superintegrable and solvable. In this case, we review the exact solutions of the corresponding Schr\"odinger equation in the Sec. III. This Section exposes a detailed comparison between two Lie-algebraic representations of the Hamiltonian as well as the explanation on the degeneracy of the system. The next Section IV treats the ground state solution at $R>0$ within the perturbative and variational formalism. In Sec. V we depict the implementation of the Lagrange Mesh Method to be used in the study of lowest excited states. For $R\ge0$ such a method leads to highly accurate results of the energies. These are displayed and discussed in Sec. VI. Here, the partial splitting of the degenerate states is established. Finally, the Section VII contains the conclusions and future work.

\section{Generalities}

\begin{figure}[h]
\centering
\includegraphics[width=8.0cm]{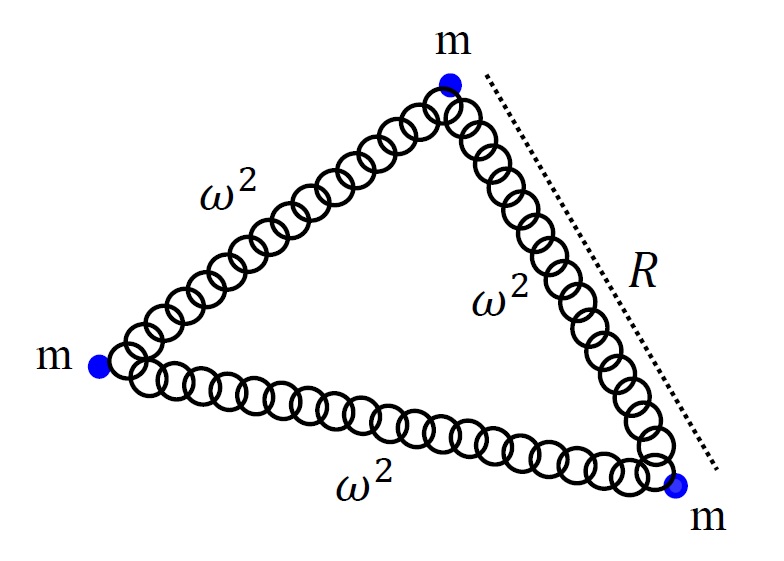} 
\caption{3-body chain of harmonic oscillators. At the minimum of the potential $V_{R}$ (\ref{Vp}), the system forms an equilateral triangle with sides $r_{12}=r_{13}=r_{23}=R$. }
\label{3-body0}
\end{figure}

Let us consider the quantum system of three non-relativistic identical particles with pairwise harmonic interaction. The potential is of the form:

 \begin{equation}
 \label{Vp}
     V_{R}\, \equiv \, \frac{3}{2} \,m\, \omega^{2} \,\left[ {(r_{12}-R)}^2 \, + \,  {(r_{13}-R)}^2  \, + \,  {(r_{23}-R)}^2 \right]\ ,
 \end{equation}
${\mathbf r}_i \in \mathbb{R}^3$, $r_{ij}=|{\mathbf r}_i-{\mathbf r}_j|$ are the relative distances between the $i$th and $j$th particles, $m$ is the common mass of each body, $\omega>0$ plays the role of angular frequency and $R$ denotes the rest length of the system, see Fig. \ref{3-body0}. The minimum of $V_{R}$ (\ref{Vp}) corresponds to an equilateral triangle with each side equal to $R$. At $R=0$, the potential $V_{R}$ is known under the name of harmonic molecule, see \cite{FF} for the case $d=1$ where the particles move on line. 

The Hamiltonian of the system is given by 
\begin{equation}
\label{H}
    {\cal H} \ = \ \frac{1}{2\,m} {\hat {\mathbf p}_{{}_1}}^2 \ + \ \frac{1}{2\,m} {\hat {\mathbf p}_{{}_2}}^2 \ + \ \frac{1}{2\,m} {\hat {\mathbf p}_{{}_3}}^2 \ + \ V_R(r_{ij}) \ ,
\end{equation}
here ${\hat {\mathbf p}_{{}_j}} = -i\,\hbar\,\nabla_j$ stands for the canonical momentum operator associated with the particle $j$. Due to translational invariance and rotational symmetry, the center of mass momentum 
\begin{equation}
{\hat {\mathbf P}}\ = \ {\hat {\mathbf p}_{{}_1}}\ + \ {\hat {\mathbf p}_{{}_2}}\ + \ {\hat {\mathbf p}_{{}_3}} \ ,
\label{pseudo}
\end{equation}
as well as the total angular momentum
\begin{equation}
{\hat{\mathbf  L}}\ = \ {\mathbf r}_1\times{\hat {\mathbf p}_{{}_1}}\ + \ {\mathbf r}_2\times{\hat {\mathbf p}_{{}_2}}\ + {\mathbf r}_3\times\ {\hat {\mathbf p}_{{}_3}} \ ,
\label{angularmom}
\end{equation}
are conserved quantities, respectively, i.e. they commute with the Hamiltonian (\ref{H}). The corresponding  stationary Schr\"odinger equation 
\begin{equation}
\label{origsp}
    {\cal H}\,\Psi({\mathbf r}_1,\,{\mathbf r}_2,\,{\mathbf r}_3) \ = \ E\,\Psi({\mathbf r}_1,\,{\mathbf r}_2,\,{\mathbf r}_3) \ ,
\end{equation}
is 9-dimensional. 

Some remarks are in order:
\begin{itemize}

\item At $R=0$, the Hamiltonian ${\cal H}$ (\ref{H}) admits a complete separation of variables in Jacobi coordinates \cite{Castro1993ExactSF}, see below. This separability holds even for the case of non-equal masses. Moreover, when all three masses are equal then the system becomes maximally superintegrable and exactly solvable \cite{Turbiner_2020}.
It is worth mentioning that for the same solvable model when restricted to a subdomain of the original configuration space, the supersymmetrization (SUSY realization) immediately encounters
subtle difficulties \cite{Znojil}.
\item For $R\neq 0$, not a singly exact solution to the Schr\"odinger equation is known so far. Interestingly, the corresponding classical system exhibits a rich dynamics with mixed regions of regularity and chaos \cite{Classicalcase}.
\end{itemize}

The integrals of motion (\ref{pseudo})-(\ref{angularmom}) allows us to construct a reduced Hamiltonian which describes all the states of ${\cal H}$ (\ref{H}) with zero total angular momentum $L =0$ ($S$-states). It solely depends on the three variables $r_{ij}$ \cite{Turbiner_2017}. Explicitly, such a reduced Hamiltonian reads
\begin{equation}
\label{Hrad}
    {\cal H}_{L=0} \ = \ -\frac{1}{2\,m}\triangle_{\rm rad} \ + \ \frac{3}{2} \,m\, \omega^{2} \,\left[ {(r_{12}-R)}^2 \, + \,  {(r_{13}-R)}^2  \, + \,  {(r_{23}-R)}^2 \right] \ ,
\end{equation}
where the kinetic-like term can be written as follows:
\begin{equation}
\begin{split}
& \frac{1}{2}\triangle_{\rm rad}\  =\   \frac{1}{r_{12}^{2}}\frac{\partial}{\partial r_{12}}\left(r_{12}^{2}\frac{\partial}{\partial r_{12}}\right)\ + \ \frac{1}{r_{13}^{2}}\frac{\partial}{\partial r_{13}}\left(r_{13}^{2}\frac{\partial}{\partial r_{13}}\right)
\ + \ \frac{1}{r_{23}^{2}}\frac{\partial}{\partial r_{23}}\left(r_{23}^{2}\frac{\partial}{\partial r_{23}}\right) \ + \
 \\ & 
  \frac{r_{12}^{2}+r_{13}^{2}-r_{23}^{2}}{2\,r_{12}\,r_{13}}\,\frac{\partial^2}{\partial r_{12}\,\partial r_{13}}
 \ + \    \frac{r_{12}^{2}+r_{23}^{2}-r_{13}^{2}}{2\,r_{12}\,r_{23}}\,\frac{\partial^2}{\partial r_{12}\,\partial r_{23}} \ + \  \frac{r_{23}^{2}+r_{13}^{2}-r_{12}^{2}}{2\,r_{23}\,r_{13}}\,\frac{\partial^2}{\partial r_{23}\,\partial r_{13}} \ ,
\end{split}
\end{equation}
with $\hbar=1$, see \cite{doi:10.1063/1.4994397} and references therein. The operator (\ref{Hrad}) describes a 3-dimensional point particle moving in a (curved) space \cite{Turbiner_2017}. For the $S-$states, the relevant spectral problem occurs in the 3-dimensional space of radial relative motion ${\mathbb{R}}_{\rm rad}=(r_{12},r_{13},r_{23})\in {\mathbb{R}}^{3}_+$
\begin{equation}
{\cal H}_{L=0}\,\psi(r_{12},r_{13},r_{23}) \ = \    E\,\psi(r_{12},r_{13},r_{23}) \ . 
\label{hred}
\end{equation}

At $R=0$, the Hamiltonian (\ref{Hrad}) is $\mathbb{Z}_2^{\otimes 3}$-invariant under the reflections 
\[
r_{12}\rightarrow -r_{12} \quad ; \qquad r_{13}\rightarrow -r_{13} \quad ; \qquad r_{23}\rightarrow -r_{23} \ ,
\]
and also under the action of the ${\mathcal S}_3^{\otimes 2}={\mathcal S}_3(r_{12},r_{13},r_{13}) \oplus {\mathcal S}_3(1,2,3)$ symmetry (permutation of the three variables $r_{12},r_{13},r_{13},$ and interchange of any pair of particles). In the case $R\neq 0$, the discrete symmetry $\mathbb{Z}_2^{\otimes 3}$ is absent. 

The Hamiltonian (\ref{Hrad}) is essentially self-adjoint with respect to the measure

\begin{equation}\label{rm}
d^3r \ = \ 8\,\pi^2\,r_{12}\,r_{13}\,r_{23}\,dr_{12}\,dr_{13}\,dr_{23} \ .
\end{equation}

\begin{figure}[h]
\centering
\includegraphics[width=7.0cm]{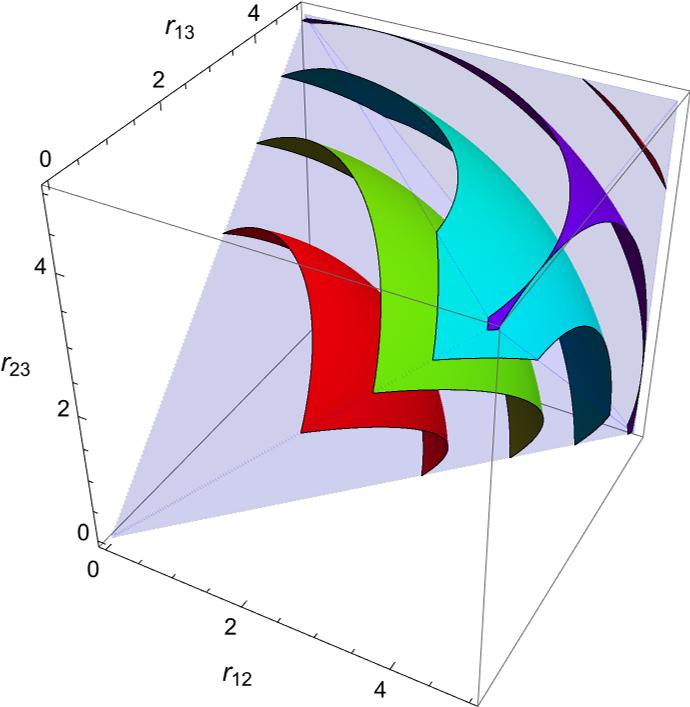} 
\caption{The domain $S\geq 0$ (shadow region), in the space of relative motion ${\mathbb{R}}_{\rm rad}=(r_{12},r_{13},r_{23})\in {\mathbb{R}}^{3}_+$, for the harmonic potential $V_{R}$ (\ref{Vp}) with $m\,\omega^2=2/3$ and unit rest length $R=1$~a.u. The other surfaces (in color) correspond to level surfaces of $V_{R}$ (equipotentials).  }
\label{cplot}
\end{figure}

Accordingly, the configuration space ${\mathbb{R}}_{\rm rad}=(r_{12},r_{13},r_{23})$ is defined by the 
inequality $S \geq 0$ (see Figure~\ref{cplot}), where $S$ is the area of the 
triangle formed by the three masses. Explicitly, using Heron's formula, the area (squared) $S^2$ reads 
\[
S^2 \ = \ \frac{1}{16} \,   (r_{12}+r_{13}+r_{23})(r_{12}+r_{13}-r_{23})(r_{12}-r_{13}+r_{23})(-r_{12}+r_{13}+r_{23})  \ .
\]

\textit{Scaling Relation}

The 3 parameters ($m,\,\omega,\,R$) in (\ref{Hrad}) define completely the harmonic 3-body system. Interestingly, one can relate
two different systems ($m,\,\omega,\,R$) and ($m',\,\omega',\,R'$) by making the scale
transformation $r_{ij} \rightarrow \sqrt{\frac{m'\,\omega'}{m\,\omega}}\,r_{ij}$. That way, the following
simple scaling relation between the two corresponding
energies holds
\begin{equation}
    E[m,\,\omega,\,R] \ = \ \frac{\omega}{\omega'}\,E\bigg[m',\,\omega',\,R'=\sqrt{\frac{m\,\omega}{m'\,\omega'}}\,R\bigg] \ .
\label{scal}
\end{equation}

\section{Case $R=0$: exact solutions}

\subsection{Jacobi-representation}

Let us consider the original Hamiltonian ${\cal H}$~(\ref{H})
which, by taking $R=0$, becomes
\begin{equation}
\label{Hjac}
    {\cal H} \ = \ \frac{1}{2\,m} {\hat {\mathbf p}_{{}_1}}^2 \ + \ \frac{1}{2\,m} {\hat {\mathbf p}_{{}_2}}^2 \ + \ \frac{1}{2\,m} {\hat {\mathbf p}_{{}_3}}^2 \ + \ \frac{3}{2} \,m\, \omega^{2} \,\left[ r_{12}^2 \, + \,  r_{13}^2  \, + \,  r_{23}^2 \right]\ .
\end{equation}
This Hamiltonian (\ref{Hjac}) admits a complete separation of variables
in Jacobi coordinates \cite{Castro1993ExactSF, Turbiner_2020, LMD:1960}. 
In the center-of-mass reference frame it takes the form
\begin{equation}
\label{H-osc-diag}
   {\cal H}\ = \
       -\frac{1}{2}\bigg[\,\frac{\partial^2}{\partial{{\bf r}_1^{(J)}} \partial{{\bf r}_1^{(J)}}} \ + \ \frac{\partial^2}{\partial{{\bf r}_2^{(J)}} \partial{{\bf r}_2^{(J)}}} \,\bigg]\ + \ \frac{9}{2}\, \omega^2\, { { ({\bf r}_1^{(J)}} } \cdot { {\bf r}_1^{(J)}}) \ + \  \frac{9}{2}\, \omega^2\, { { ({\bf r}_2^{(J)}} } \cdot { {\bf r}_2^{(J)}}) \ ,
\end{equation}
($\hbar=1$ and $m=1$) where
\begin{displaymath}
\label{Jacobi}
{\bf r}^{(J)}_{1} \ = \ \sqrt{\frac{1}{2}}\left( {\bf r}_{1}\ -\ {\bf r}_{2}\right), \qquad
{\bf r}^{(J)}_{2} \ = \ \sqrt{\frac{2}{3}}\left({\bf r}_{3}\ -\ \frac{{\bf r}_{1}+{\bf r}_{2}}{2}\right)
\end{displaymath}
are nothing but the two 3-dimensional vector Jacobi coordinates. In these variables, 
the Hamiltonian (\ref{Hjac}) corresponds to the sum of two identical 3-dimensional isotropic harmonic oscillators, more precisely, of two \textit{Jacobi harmonic oscillators} \cite{Turbiner_2020}. Hence, the system is maximally superintegrable and exactly solvable. In particular, in this Jacobi-representation the second quantization formalism in terms of creation and annihilation operators, namely
\begin{equation}
{\cal H} \ = \ \sum_{\alpha} a^{\dagger}_\alpha\,a_\alpha \ + \ 9\,\omega     \ ,
\end{equation}
($\alpha$ is an index of a number of quantum numbers) can be introduced immediately. As demonstrated in \cite{From2dimensional}, there exists a connection between the theory of superintegrable 2-dimensional
systems (on the plane) and the theory of superintegrable systems (on the space of relative
motion) of the 3-body problem parameterized by Jacobi distances. The spectra $E$ of (\ref{Hjac}) is the sum of spectra of individual Jacobi oscillators
\begin{equation}
\label{EJ}
E \ = \ \Lambda_1 \ + \ \Lambda_2 \ \equiv \ 3\,\omega\,\big(\, 2\,[n_1\,+\,n_2] \ + \ell_1 \ + \ \ell_2   \ + \ 3\,\big) \ ,
\end{equation}
and its eigenfunctions are the product 
\begin{equation}
\label{eigJ}
\Psi( {\bf r}^{(J)}_{1},\, {\bf r}^{(J)}_{2})\ = \ \Psi_1^{(J)}( {\bf r}^{(J)}_{1})\times\Psi_2^{(J)}( {\bf r}^{(J)}_{2})\ ,
\end{equation} 
of the well-known individual solutions, see below. In spherical coordinates, 
${\bf r}^{(J)}_{k} \equiv ( r^{(J)}_{k},\,\theta^{(J)}_k,\,\phi^{(J)}_k)$, they are given by
\begin{equation}
\label{psij}
\Psi_k^{(J)}( {\bf r}^{(J)}_{k};\,n_k,\,\ell_k) \ = \ {\tau_k}^{\frac{\ell_k}{2}}\,e^{-\frac{3\,\omega}{2}\,  \tau_k} \,L_{n_k}^{\ell_k+\frac{1}{2}}(3\,\omega\, \tau_k)  \,\,Y_{\ell_k,\,s_k}(\theta^{(J)}_k,\,\phi^{(J)}_k)\,, \qquad k=1,2 \ ,
\end{equation}
where $\tau_k={( r^{(J)}_{k})}^{2}$, $\,L_{n_k}^{\ell_k+\frac{1}{2}}(3\,\omega\, \tau_k)$ is a generalized Laguerre polynomial and $Y_{\ell_k,\,s_k}$ denotes a spherical harmonic function. In (\ref{EJ})-(\ref{psij}), $n_k$ and $\ell_k$ are the individual radial and angular-momentum quantum numbers, respectively, of the $k$th Jacobi
oscillator. In this representation the eigenfunctions of the Hamiltonian (\ref{H-osc-diag}) are labeled by six quantum integers numbers ($n_1,\,\ell_1,\,s_1;\,n_2,\,\ell_2,\,s_2$).

In the subspace of fixed $\ell_1$ and $\ell_2$, the Hamiltonian (\ref{H-osc-diag}) reduces to
\begin{equation}
\label{}
   {\cal H}\ = \
       -\frac{1}{2}\sum_{k=1}^2\bigg[\,\frac{\partial^2}{\partial{{r}_k^{(J)}} \partial{{ r}_k^{(J)}}} \ + \ \frac{2}{{r}_k^{(J)}}\frac{\partial}{\partial{{r}_k^{(J)}} } \ + \ \frac{\ell_k\,(\ell_k+1)}{{\big({r}_k^{(J)}\big)}^2} \,\bigg]\ + \ \frac{9}{2}\, \omega^2\,\big[\,{\big({r}_1^{(J)}\big)}^2 \ + \ {\big({r}_2^{(J)}\big)}^2\,\big] \ .
\end{equation}
It acts on the $2$-dimensional space (${r}_1^{(J)},\,{r}_2^{(J)}$) of radial 
Jacobi distances alone. Evidently, it possesses a hidden Lie algebra $sl_2^{\,\otimes \,{2}}$. 

Moreover, for fixed $\ell_1$ and $\ell_2$ the zero total angular momentum solutions 
($S$-states) of the system are characterized by $\ell_1=\ell_2\equiv \ell$, $s_1=-s_2$ 
with $n_1,n_2$ arbitrary. Otherwise in (\ref{H-osc-diag}) the sum of the individual 
angular momentum  of the two Jacobi oscillators is always different from zero. In the 
case $\ell_1=\ell_2\equiv \ell$ with $s_1=s_2=0$ it can be shown that the degeneracy 
of the $N$th-level $E=3\,\omega\,(2N+3)$, with $N= n_1+n_2+\ell $, is $g=\frac{(N+1)(N+2)}{2}$.

\subsection{$\rho$-representation}

If we now consider the reduced Hamiltonian ${\cal H}_{L=0}$ in (\ref{hred}),
it is  found that at $R=0$ it admits special solutions in the factorized form
\begin{equation}
\label{psiN}
\psi^{(g)}_{N} \ = \ P_{N,g}(r_{12}^2,\,r_{13}^2,\,r_{23}^2)\times e^{-\frac{\omega}{2}(r_{12}^2+r_{13}^2+r_{23}^2)} \ ,
\end{equation}
where $P_{N,g}$ is a multivariate polynomial of degree $N$ in variables 
\begin{equation}
\rho_{ij}\ = \ r_{ij}^2\,, \qquad \ i\neq j\,=\,1,2,3\ ,
\end{equation}
and $g=(N+1)(N+2)/2$ is the degeneracy of the $N$th-level $E_{N}$
\begin{equation}
\label{EN}
E_{N} \ = \ 3\,\omega\,\big(\,2\,N \ + \ 3\,\big)\,,   \qquad N=0,1,2,3\ldots \ .
\end{equation}
In these variables $\rho_{ij} =r_{ij}^2$, the Hamiltonian ${\cal H}_{L=0}$ in 
(\ref{hred}) possesses a hidden Lie algebra $sl_4$, and it can be rewritten 
(up to a gauge transformation) in terms of $sl_4$ algebra generators.
The eigenvalues become linear in 3 quantum numbers \cite{Turbiner_2020}, 
namely $N=N_1+N_2+N_3$. We emphasize that (\ref{hred}) does not separate in 
the $\rho-$variables. 

In particular, for the normalized ground state $N=0$ eigenfunction we obtain 
\begin{equation}
\label{psiN0}
\psi_{0} \ = \ \frac{2\, \pi}{3 \,\sqrt{3}\,  \omega^{3} }   e^{-\frac{\omega}{2}(\,\rho_{12}+\rho_{13}+\rho_{23}\,)}\ ,
\end{equation}
with energy $E_0= 9\,\omega$, whereas for the first excited state $N=1$ there exist 3 degenerate solutions of the form (\ref{psiN}), explicitly they read
\begin{equation}
\begin{split}
    &      \psi_{1}^{(0)}\ = \  \frac{(3)^{\frac{5}{4}}\omega^{\frac{3}{2}}}{\sqrt{2\,\pi}} \,(\,1 \ - \ \omega\, \rho_{12})\,e^{-\frac{\omega}{2}  (\,\rho_{12}+\rho_{13}+\rho_{23}\,)}
\\ &
\psi_{1}^{(1)}\ = \  \frac{(3)^{\frac{5}{4}}\omega^{\frac{3}{2}}}{\sqrt{2\,\pi}} \,(\,1 \ - \ \omega\, \rho_{13})\,e^{-\frac{\omega}{2}  (\,\rho_{12}+\rho_{13}+\rho_{23}\,)}
\\ &
 \psi_{1}^{(2)}\ = \  \frac{(3)^{\frac{5}{4}}\omega^{\frac{3}{2}}}{\sqrt{2\,\pi}} \,(\,1 \ - \ \omega\, \rho_{23})\,e^{-\frac{\omega}{2}  (\,\rho_{12}+\rho_{13}+\rho_{23}\,)} \ ,
\end{split}    
\end{equation}

corresponding to the same energy $E_1=15\,\omega$.

\bigskip

In this way, in the case $R=0$ we arrive to the following results:

\begin{itemize}
\item For the $S-$states of (\ref{H-osc-diag}), the subset of the exact eigenfunctions 
$\Psi( {\bf r}^{(J)}_{1},\, {\bf r}^{(J)}_{2})$ (\ref{eigJ}) in 
Jacobi coordinates characterized by $\ell_1=\ell_2= \ell$ and 
$s_1=s_2=0$ can be completely expressed as a linear combination of 
the $\rho$-dependable special solutions (\ref{psiN}). To illustrate this point, 
for the lowest states $N=0,1,2$ we indicate, see Table \ref{TDeg}, 
the degeneration of the system in both representations.

\vspace{0.15cm}
\item The ground state eigenfunction of the original Hamiltonian (\ref{H}) solely depends on the three $\rho-$variables, $\rho_{ij}=r^2_{ij}$, the three relative distances (squared). It is presented in (\ref{psiN0}).

\vspace{0.15cm}

\item For the $S-$states of (\ref{H-osc-diag}), the exact eigenfunctions 
$\Psi( {\bf r}^{(J)}_{1},\, {\bf r}^{(J)}_{2})$ (\ref{eigJ}) in 
Jacobi coordinates characterized by $\ell_1=\ell_2= \ell$ but 
$s_1\neq s_2$ can not be expressed as a polynomial in $\rho-$variables multiplied by the ground state eigenfunction.

\end{itemize}

\begin{center}
\begin{table}[h]
\caption{$S-$states: Degeneration of the energy level 
$E_N=3\,\omega\,(\,2\,N \, + \, 3\,)$ in (I) the Jacobi-representation 
($N=n_1+n_2+\ell$) with $s_1=s_2=0$ vs 
(II) the $\rho-$representation ($N=N_1+N_2+N_3$).}
\begin{tabular}{l | ccc|ccc}
\hline\hline
$N$  \hspace{0.1cm}    &  \hspace{0.1cm} $n_1$    \hspace{0.1cm}   & 
$n_2$\hspace{0.1cm}  & \hspace{0.1cm}$\ell$ \hspace{0.1cm} & 
          \hspace{0.1cm} $N_1$  \hspace{0.1cm} & 
\hspace{0.1cm} $N_2$  \hspace{0.1cm} &  \hspace{0.1cm} $N_3$  \hspace{0.1cm} \\
\hline
0& 0 & 0 & 0 & 0 & 0 & 0 \\
\hline\hline
1& 1 & 0 & 0 & 1 & 0 & 0 \\  
1& 0 & 1 & 0 & 0 & 1 & 0 \\
1& 0 & 0 & 1 & 0 & 0 & 1 \\
\hline\hline
2& 2 & 0 & 0 & 2 & 0 & 0 \\  
2& 0 & 2 & 0 & 0 & 2 & 0 \\
2& 1 & 1 & 0 & 0 & 0 & 2 \\
2& 1 & 0 & 1 & 1 & 1 & 0 \\  
2& 0 & 1 & 1 & 1 & 0 & 1 \\
2& 0 & 0 & 2 & 0 & 1 & 1 \\
\hline\hline
\end{tabular}
\label{TDeg}
\end{table}
\end{center}
It will be shown that for $R>0$ and fixed $N$, the original degenerate $g-$eigenfunctions with $g=\frac{(N+1)(N+2)}{2}$ split into $(N(N+1)+2)/2$ different energy sub-levels.  

\section{Case $R>0$: ground state}

In this Section, as a first step we study the ground state of the spectral problem (\ref{hred}) as a function of the rest length $R$. To find the corresponding approximate solutions, the perturbative and the variational method are employed.

\subsection{Small $R$: perturbation theory}

Taking the $R-$dependent terms in the potential $V_R$ (\ref{Vp}) as a perturbation to the exactly solvable problem $V_{R=0}$, one can immediately compute the first correction to the energy within the non-linearization procedure \cite{NLP}. Putting $m=1$, the original potential $V_R$~(\ref{Vp}) can be written as a sum of three terms
\begin{equation}
\label{VRsplit}
 V_{R}\, = \, \frac{3}{2}  \,\omega^{2} \,\left[ r_{12}^2 \, + \,  r_{13}^2  \, + \,  r_{23}^2\right] \ - \ 3\,\omega^2\,R\,(r_{12}+r_{13}+r_{23}) \ + \ \frac{9}{2}\,\omega^2\,R^2  \ ,
\end{equation}
where the first term corresponds to an exactly solvable potential whereas the last one is just a constant. Let us write the ground state function in exponential form:
\[
\psi_0(r_{12},\,r_{13},\,r_{23}\,) \ = \ e^{-\Phi_0(r_{12},\,r_{13},\,r_{23}\,)} \ ,
\]
here $\Phi$ is the phase of the wave function. Next, we develop perturbation theory in powers of $R$, namely
\begin{equation}
\begin{aligned}
E_0 \ = \  \sum_{n=0}^{\infty}\epsilon_n\,R^n\qquad ; \qquad \Phi_0(\,r_{12},\,r_{13},\,r_{23}\,) \ = \ \sum_{n=0}^{\infty}a_n(r_{12},\,r_{13},\,r_{23}\,)\,R^n \ ,
\end{aligned}    
\end{equation}
with $\epsilon_0=9\,\omega$ and $a_0={\frac{\omega}{2}(r_{12}^2+r_{13}^2+r_{23}^2)}$ being the exact ground- state solution occurring for $R=0$. As a result of direct calculations, we obtain the value 
\begin{equation}
\label{E0PT1}
    E_{0}(R)\ = \  9\,\omega \ + \  \frac{3\,\omega}{2\,\pi}(\,3\,\pi\, \omega\,R^{2} \ - \ 4\,R\,\sqrt{6\,\pi\, \omega}\,) \ + \ {\rm corrections} \ .
\end{equation}

Therefore, taking the zero and first order corrections as well as the constant term appearing in (\ref{VRsplit}), perturbation theory predicts the existence of a global minimum in $E_0(R)$, localized at $R=R_{\rm crit} \sim 2\sqrt{\frac{2}{3\,\pi\,\omega}}$, which would corresponds to an equilateral configuration of equilibrium. Since this phenomenon could be an artifact of the deficiency of perturbation theory we will verify it using different approximate methods, see below.

\subsection{Arbitrary $R$: variational method}

For arbitrary $R$, to evaluate the ground state energy $E_0(R)$ we take the simple two-parametric trial function 
\begin{equation}
\label{psitrial}
    \psi_{\rm trial} \ = \ e^{-\frac{\alpha\,\omega}{2}\big(\,{(r_{12}\,-\,\beta\,R)}^2\ + \ {(r_{13}\,-\,\beta\,R)}^2\ + \ {(r_{23}\,-\,\beta\,R)}^2\,\big)} \ ,
\end{equation}
where $\alpha,\,\beta \in [0,\,1]$ are variational parameters. At $\beta=0$ 
with $\alpha=1$, the function (\ref{psitrial}) degenerates into the 
exact solution appearing at $R=0$. Though the value of the minimum 
in $E_0(R)$ disagrees with that found in perturbation theory, the 
variational method also predicts the existence of a global minimum. 
At fixed $\omega=1$, a comparison of the values for $E_0(R)$ obtained 
with the perturbative (\ref{E0PT1}), variational  (using the optimal 
(\ref{psitrial})) and Lagrange-mesh method (see below), respectively, 
is displayed in Figure~\ref{G1} within the interval $ R \in [0,\,3.5]$\,a.u. 
In this case, perturbation theory (\ref{E0PT1}) provides reasonable 
accurate results when $0 \leq R \leq 0.3$\,a.u. whereas for the variational method 
the corresponding interval of applicability is larger $0 \leq R \leq 2.3$\,a.u., 
as expected. 
A quantitative comparison between the variational  and Lagrange-mesh 
results is presented in Table~\ref{TabVM}.
In particular, the relative difference increases from $0.003\%$ at $R=0.2$~a.u. 
up to $9\%$ at $R=7/2$~a.u.

\begin{figure}[h]
\centering
\includegraphics[width=10.0cm]{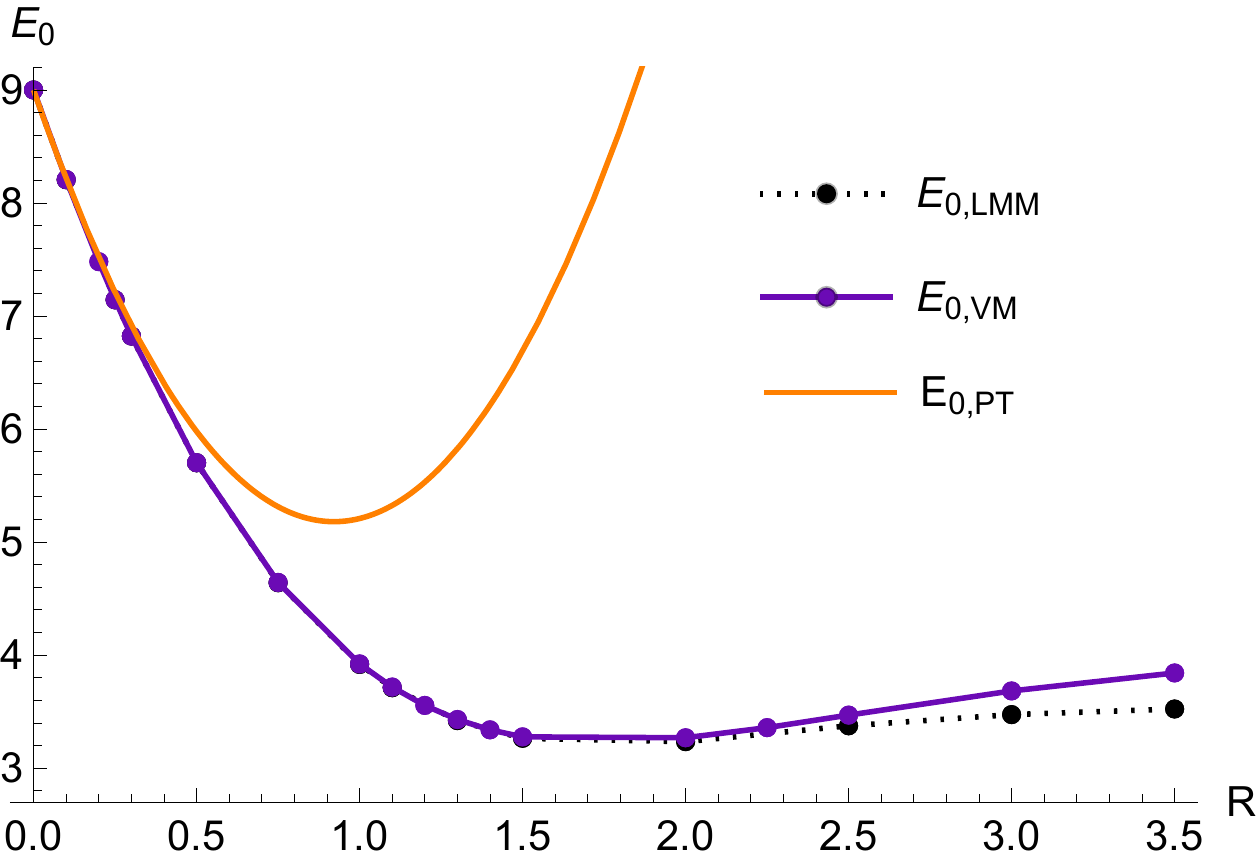} 
\caption{Comparison of the ground state energy $E_0$ in a.u. calculated using 
Perturbation Theory (PT)~(\ref{E0PT1}), the variational function $\psi_{\rm trial}$ 
(VM)~(\ref{psitrial}) and the Lagrange-Mesh Method (LMM), see below, at $\omega=1$. }
\label{G1}
\end{figure}

\begin{center}
\begin{table}[h]
\caption{Variational ground state energy $E_0^{\rm{VM}}$ vs $R$ computed using 
the 2-parametric trial function $\psi_{\rm trial}$ (\ref{psitrial}) 
with $\omega=1$.  The optimal values of the parameters $\alpha$ and $\beta$ 
are also displayed. $E_{0}^{\rm{LMM}}$ corresponds to the value of the 
energy calculated using the Lagrange-mesh method (see below). The relative 
error (RE) is presented in the last column.}
\begin{tabular}{|c||l|ll||c||l|}
\hline\hline
$R$&\multicolumn{1}{c|}{$E_0^{\rm{VM}}$} &\multicolumn{1}{c}{$\alpha$} 
   &\multicolumn{1}{c||}{$\beta$}
& $E_{0}^{\rm{LMM}}$&\multicolumn{1}{c|}{RE}\\ 
\hline
0.0 &\qquad 9      &\quad 1    & \quad 0    & 9.000000000000&\quad\, 0      \\
0.2 & 7.48081& 0.960& 0.300& 7.480575209823& 0.00003\\
0.3 & 6.82304& 0.935& 0.300& 6.822630580417& 0.00006\\
0.5 & 5.70206& 0.875& 0.310& 5.700976066954& 0.00019\\
1.0 & 3.92165& 0.790& 0.440& 3.916842671126& 0.00123\\
1.1 & 3.71684& 0.770& 0.460& 3.710814446829& 0.00162\\
1.2 & 3.55424& 0.760& 0.500& 3.546619661941& 0.00215\\
1.3 & 3.42955& 0.760& 0.544& 3.420165435764& 0.00274\\
1.4 & 3.33804& 0.760& 0.577& 3.327196880347& 0.00326\\
1.5 & 3.27597& 0.760& 0.610& 3.263367579471& 0.00386\\
2.0 & 3.26821& 0.754& 0.730& 3.233525277971& 0.01073\\
2.5 & 3.46849& 0.750& 0.843& 3.373848351684& 0.02805\\
3.0 & 3.68241& 0.750& 0.900& 3.473009053896& 0.06029\\
3.5 & 3.84142& 0.750& 0.930& 3.522515737857& 0.09053\\
\hline\hline
\end{tabular}
\label{TabVM}
\end{table}
\end{center}

\section{The Lagrange-mesh method}
In order to solve the Schr\"odinger equation for the 
Hamiltonian~(\ref{Hrad}), the Lagrange-mesh method (LMM) is also 
applied~\cite{BH:1986,DB:2015}. To introduce this methodology 
let's consider a one-dimensional problem where
a set of $M$ Lagrange functions $f_i\,(x)$ defined over the  
domain $x\in[\,0,\infty)$ is associated with $M$ mesh points $x_i$ 
which correspond to the zeros of Laguerre polynomials of degree $M$,
{\it i.e.} $L_M(x_i) = 0$. The Lagrange-Laguerre functions 
$f_i\,(x)$ which satisfy the Lagrange conditions 
\begin{equation}
\label{lagcond}
f_i(x_j)=\lambda_i^{-1/2}\delta_{ij}\,,
\end{equation}
at the $M$ mesh points  are given by
\begin{equation}
\label{lagF}
f_i(x)=(-1)^{i} \frac{x}{x_i^{1/2}}\frac{L_M(x)}{(x-x_i)}e^{-x/2}\,,
\end{equation}
and the coefficients $\lambda_i$ are the weights associated with a 
Gauss quadrature 
\begin{equation}
\label{gqcond}
\int_0^{\infty} G(x) dx \approx \sum_{k=1}^{M} \lambda_k\, G(x_k)\,.
\end{equation}
In terms of the $M$  Lagrange functions $f_i(x)$~(\ref{lagF}), the
solution of the Schr\"odinger equation for a particle of mass $m$ 
in a potential $V(x)$ is expressed as 
\begin{equation}
\psi(r) =\sum_{i=1}^{M}c_i f_i(r)\,.
\label{lmfunc}
\end{equation}
The function~(\ref{lmfunc}), together with the Gauss 
quadrature~(\ref{gqcond}) and the Lagrange 
conditions~(\ref{lagcond}) leads to the system of 
variational equations
\begin{equation}
\label{lmmeqs}
\sum_{j=1}^M \left[\frac{\hbar^2}{2m}\,T_{ij}+
             V(x_i)\delta_{ij}\right]\,c_j = E\,c_i\,,
\end{equation}
where $T_{ij}$ are the kinetic-energy matrix elements 
(see for example~\cite{DB:2015}) and  $V(x_i)$ is the potential 
evaluated at the mesh points $x_i$. By solving the 
system~(\ref{lmmeqs}), the energies $E$ and the eigenvectors 
$c_{i}$ are obtained, from which the approximation to the wave 
function~(\ref{lmfunc}) is obtained.

The present system has three degrees of freedom which are 
described by the three distances between the particles: 
$r_{12}$, $r_{13}$ and $r_{23}$. 
Technically, a considerable simplification results from going
over to the so-called perimetric coordinates used by Pekeris 
in his helium calculations \cite{Pekeris1958}. These perimetric 
coordinates are defined by the linear relations
\begin{equation}
\begin{aligned}
& x  \ = \ r_{12} \ + \ r_{13} \ - \ r_{23}\,,\\ 
& y  \ = \ r_{12} \ - \ r_{13} \ + \ r_{23}\,,\\
& z  \ = \ -r_{12} \ + \ r_{13} \ + \ r_{23}\,.
\end{aligned}
\label{perimetric}
\end{equation}
The volume element is $dV \propto (x+y)(x+z)(y+z)dx\,dy\,dz$.
By the above transformation the limits of the three perimetric
coordinates $x$, $y$ and $z$ become independent of each other 
and they vary from $0$ to $\infty$.

In perimetric coordinates, following the notation presented 
in~\cite{BJ:2015}, the matrix elements of the kinetic energy 
operator~(\ref{Hrad}) $\langle F |\hat{T}|G \rangle$ between 
functions $F$ and $G$ can be written as
\begin{equation}
\langle F |\hat{T}|G \rangle = \int_{0}^{\infty}\,dx\,
\int_{0}^{\infty}\,dy\,\int_{0}^{\infty}\,dz \sum_{i,j=1}^{3} A_{ij}(x,y,z) \pder{F}{x_i}\pder{G}{x_j}\,,
\label{mtxFG}
\end{equation}
where $(x_1,x_2,x_3)=(x,y,z)$ and the coefficients $A_{ij}$ are given by
\begin{eqnarray}
A_{1,1}&=&x(y+z)(x+y+z)+xz(x+z)+xy(x+y)\,,\\
A_{2,2}&=&yz(y+z)+y(x+z)(x+y+z)+xy(x+y)\,,\nonumber\\
A_{3,3}&=&yz(y+z)+xz(x+z)+z(x+y)(x+y+z)\,,\nonumber\\
A_{1,2}&=&A_{2,1}=-xy(x+y)\,,\nonumber\\
A_{1,3}&=&A_{3,1}=-xz(x+z)\,,\nonumber\\
A_{2,3}&=&A_{3,2}=-yz(y+z)\,.\nonumber
\label{aijcoeff}
\end{eqnarray}

The generalization of the Lagrange-mesh method to the three-dimensional 
case is as follows~\cite{HB:2001,HB:2003}. The three-dimensional Lagrange 
functions $F_{ijk}(x,y,z)$ are defined as
\begin{equation}
F_{ijk}(x,y,z)= {\cal N}_{ijk}^{-1/2}f_i^{M_x}(x/h_x)f_j^{M_y}(y/h_y)f_k^{M_z}(z/h_z)\,,
\label{3dbs}
\end{equation}
where the functions $f_i^M$ all have the same structure 
as~(\ref{lagF}) with $M$ replaced by the respective degrees 
$M_x$, $M_y$ and $M_z$ and 
the zeros $x_p$ ($p = 1,...,M_x$), $y_q$ ($q = 1,...,M_y$) and 
$z_r$ ($r = 1,...,M_z$) are the zeros of the respective 
Laguerre polynomials. The scaling parameters $h_x$, $h_y$ and 
$h_z$ are incorporated to fit the mesh to the physical system.
The normalization factor ${\cal N}_{ijk}$ is defined by
\begin{equation}
{\cal N}_{ijk}=h_x\,h_y\,h_z\,(h_x\,x_i+h_y\,y_j)(h_x\,x_i+h_z\,z_k)(h_y\,y_j+h_z\,z_k)\,.    
\end{equation}
This three-dimensional Lagrange functions~(\ref{3dbs}) satisfy
\begin{equation}
F_{ijk}(h_x\,x_{i'}, h_y\,y_{j'},h_z\,x_{k'})=({\cal N}_{ijk}\,\lambda_i\,\mu_j\,\nu_k)^{-1/2}\,\delta_{ii'}\delta_{jj'}\delta_{kk'}\,,   
\label{condL3d}
\end{equation}
where $\lambda_i$, $\mu_j$ and $\nu_k$ are the weights of the Gauss 
quadratures~(\ref{gqcond}) for the variables $x$, $y$ and $z$, respectively. 
In terms of the Lagrange functions $F_{ijk}(x,y,z)$~(\ref{3dbs}), 
the wave function is expanded as
\begin{equation}
\Psi(x,y,z)= \sum_{i=1}^{M_x}\sum_{j=1}^{M_y}\sum_{k=1}^{M_z} C_{ijk} F_{ijk}(x,y,z)\,, 
\label{wfper}
\end{equation}
which makes it possible, together with the Gauss quadratures for each 
variable and condition~(\ref{condL3d}), to write the Schrodinger equation 
for the  Hamiltonian~(\ref{Hrad}) as a mesh equation
\begin{equation}
\sum_{i=1}^{M_x}\sum_{j=1}^{M_y}\sum_{k=1}^{M_z}\left\{\langle F_{i'j'k'}|T|F_{ijk}\rangle + [\langle F_{i'j'k'} | V |F_{ijk}\rangle-E]\delta_{ii'}\delta_{jj'}\delta_{kk'}\right\} C_{ijk} =0\,.
\end{equation}
The matrix elements of the potential $\langle F_{i'j'k'} | V |F_{ijk}\rangle$ 
have a very simple representation
\begin{equation}
\langle F_{i'j'k'} | V |F_{ijk}\rangle = V(h_x\,x_i, h_y\,y_i, h_z\,z_i)\,\delta_{ii'}\,\delta_{jj'}\,\delta_{kk'}\,,
\label{lmeq}
\end{equation}
which correspond to the potential $V_R$~(\ref{Vp}) in perimetric 
coordinates evaluated at the mesh points.
In contrast, the matrix elements of the kinetic energy 
operator $\langle F_{i'j'k'}|T|F_{ijk}\rangle$ between two 
elements $F_{ijk}$~(\ref{mtxFG})  are given by
\begin{eqnarray}
\langle F_{i'j'k'} | T |F_{ijk}\rangle &=& 2\,{\cal N}_{i'j'k'}^{-1/2}\,{\cal N}_{ijk}^{-1/2}\,h_x\,h_y\,h_z\,\left\{ \right.\\
& & \delta_{jj'}\,\delta_{kk'}\,\sum_n \lambda_n\,h_x^{-2}\,A_{11}(h_x x_n,h_y y_j, h_z z_k)\,f'_{i}(x_n)\,f'_{i'}(x_n)\nonumber\\
&+& \delta_{ii'}\,\delta_{kk'}\,\sum_n \mu_n\,h_y^{-2}    \,A_{22}(h_x x_i,h_y y_n, h_z z_k)\,f'_{j}(y_n)\,f'_{j'}(y_n)\nonumber\\
&+& \delta_{ii'}\,\delta_{jj'}\,\sum_n \nu_n\,h_z^{-2}    \,A_{33}(h_x x_i,h_y y_j, h_z z_n)\,f'_{k}(z_n)\,f'_{k'}(z_n)\nonumber\\
&+& \delta_{kk'}(h_x\,h_y)^{-1}[ (\lambda_{i}\mu_{j'})^{1/2}\,A_{12}(h_x\,x_{i},h_y\,y_{j'}, h_z\, z_{k})\, f'_{i'}(x_{i})\,f'_{j}(y_{j'}) \nonumber\\
&& \hspace{2.1cm}+(\lambda_{i'}\mu_{j})^{1/2}\,A_{12}(h_x\,x_{i'},h_y\,y_{j}, h_z\, z_{k})\, f'_{i}(x_{i'})\,f'_{j'}(y_{j})]\nonumber\\
&+& \delta_{jj'}(h_x\,h_z)^{-1}[ (\lambda_{i}\nu_{k'})^{1/2}\,A_{13}(h_x\,x_{i},h_y\,y_{j}, h_z\, z_{k'})\, f'_{i'}(x_{i})\,f'_{k}(z_{k'}) \nonumber\\
&& \hspace{2.0cm}+(\lambda_{i'}\nu_{k})^{1/2}\,A_{13}(h_x\,x_{i'},h_y\,y_{j}, h_z\, z_{k})\, f'_{i}(x_{i'})\,f'_{k'}(z_{k})]\nonumber\\
&+& \delta_{ii'}(h_y\,h_z)^{-1}[ (\mu_{j}\nu_{k'})^{1/2}    \,A_{23}(h_x\,x_{i},h_y\,y_{j}, h_z\, z_{k'})\, f'_{j'}(y_{j})\,f'_{k}(z_{k'}) \nonumber\\
&& \hspace{2.0cm}+(\mu_{j'}\nu_{k})^{1/2}    \,A_{23}(h_x\,x_{i},h_y\,y_{j'}, h_z\, z_{k})\, f'_{j}(y_{j'})\,f'_{k'}(z_{k})]\nonumber
\left.\right\}\,,
\end{eqnarray}
with the coefficients $A_{ij}$ defined by~(\ref{aijcoeff}) and
$f'_i(x_j)=d f_i(x)/dx|_{x=x_j}$ and analogously for $y$ and $z$.
To obtain the solution of the mesh equation~(\ref{lmeq}),  we consider 
$M_x=M_y=M_z=M$ and $h_x=h_y=h_z=h$. The so obtained results are presented 
in the following section. 

\section{Results and discussion}

In this Section the energies $E=E(R)$ appearing in (\ref{hred}) 
for the lowest $S$-states solutions 
$\psi(x,y,z)$~(\ref{wfper}) using the Lagrange-mesh 
method are presented. 
For clarity of the degeneracy as a function of $R$, the 
energies are denoted by $E=E_{N,n}$. At $R=0$~a.u. the label $N$ corresponds 
to the quantum number of the exact solution (\ref{psiN}). For $R>0$, the 
degenerate $N$th-level splits into sub-levels denoted by $n=0,1,2,3,\ldots$.   

\begin{figure}[h]
\centering
\includegraphics[width=13.0cm]{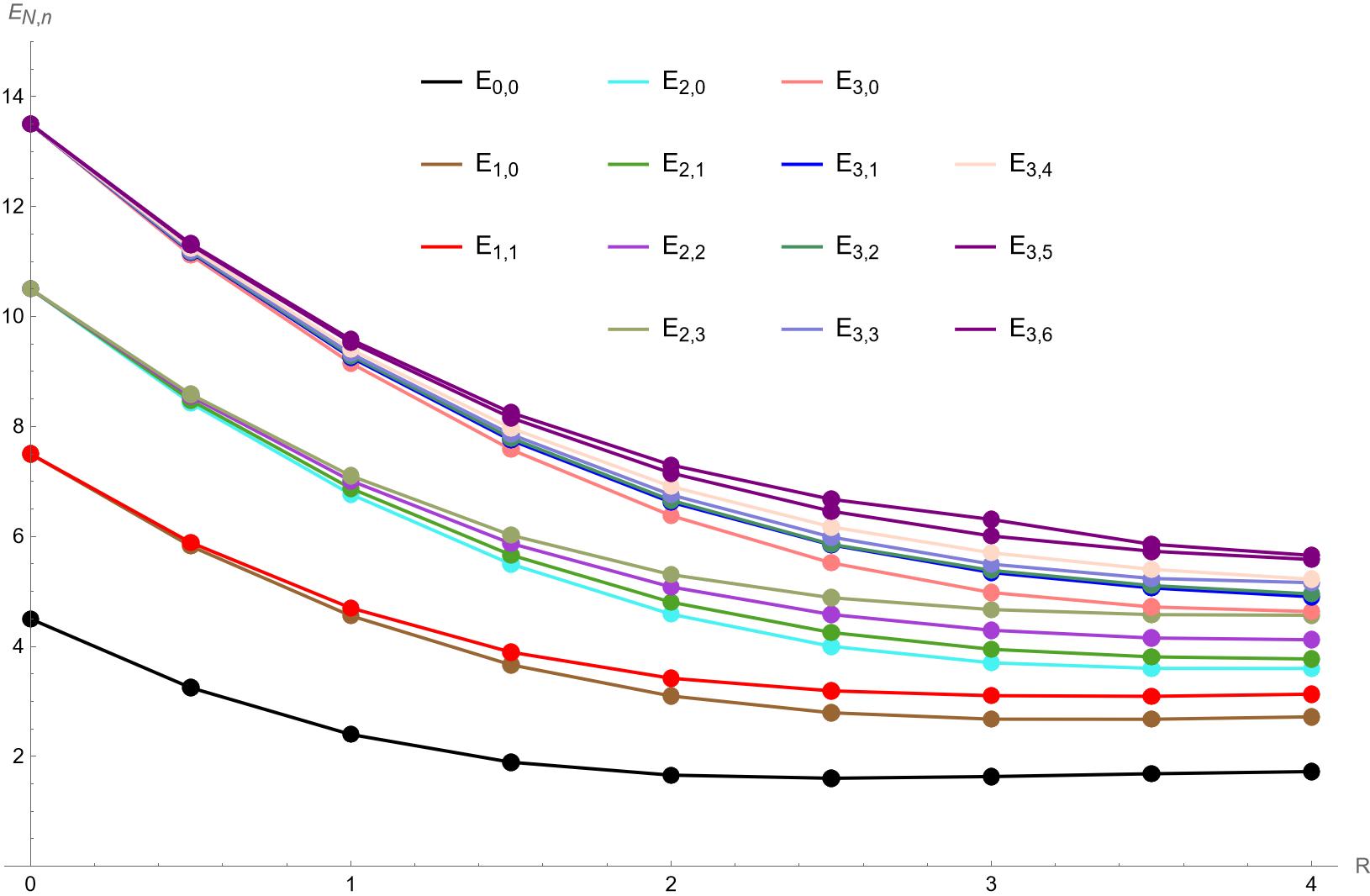}
\caption{ 
The energy $E_{N,n}$ vs $R$ of the three-body harmonic system $V_R$~(\ref{Vp}) 
for the 14 lowest $S$-states. It is a smooth function of $R$. Numerical values 
calculated with the Lagrange-Mesh method are marked by bullets, the lines are 
a guide to the eye. The values $m=1$ and $\omega=1/2$ were used in the 
calculations (color online). Energy is presented in Hartrees.}
\label{fENnw05}
\end{figure}

Table~\ref{t3bw05} presents the energy values for $\omega=0.5$ and 
$R\in [0.0,4.0]$~a.u. in constant steps of $0.5$~a.u. for four different values of $N$: 
$N=0, 1, 2$ and $3$. In all cases 12 decimal digits are provided.
The limit case $R=0$, presented in the second column, is in complete 
agreement with the analytic solution~(\ref{EN}). It can also be noticed 
that the degeneracy obtained for each $N$ value
$g_0=1$, $g_1=3$, $g_2=6$ and $g_3=10$ for $N=0, 1, 2$ 
and $3$, respectively, coincides with the values of the analytic
expression $g_N=(N+1)(N+2)/2$. For $R\ne 0$, the degeneration is partially 
removed: each $g_N$-degenerated energetic level $E_{N}$ seems to unfold into 
$(N(N+1)+2)/2$ different energy levels $E_{N,n}$, as can be seen in columns $3$ 
through $6$ of Table~\ref{t3bw05} (and its continuation) and Figure~\ref{fENnw05}
where the results are depicted.
As pointed out in the previous sections for the case $\omega=1.0$ (see also below), 
the ground state $E_{0,0}$ presents a minimum. For $\omega=0.5$ the equilibrium length 
for which the minimum appears is $R_{min}\approx 2.494583$~a.u. and the 
corresponding energy value  is $E_{0,0}^{min}=1.601390190475$~a.u. 
A global minimum in energy as a function of $R$ is present for excited 
states $E_{N,n}$ as well.

\begin{figure}[h]
\centering
\includegraphics[width=13.0cm]{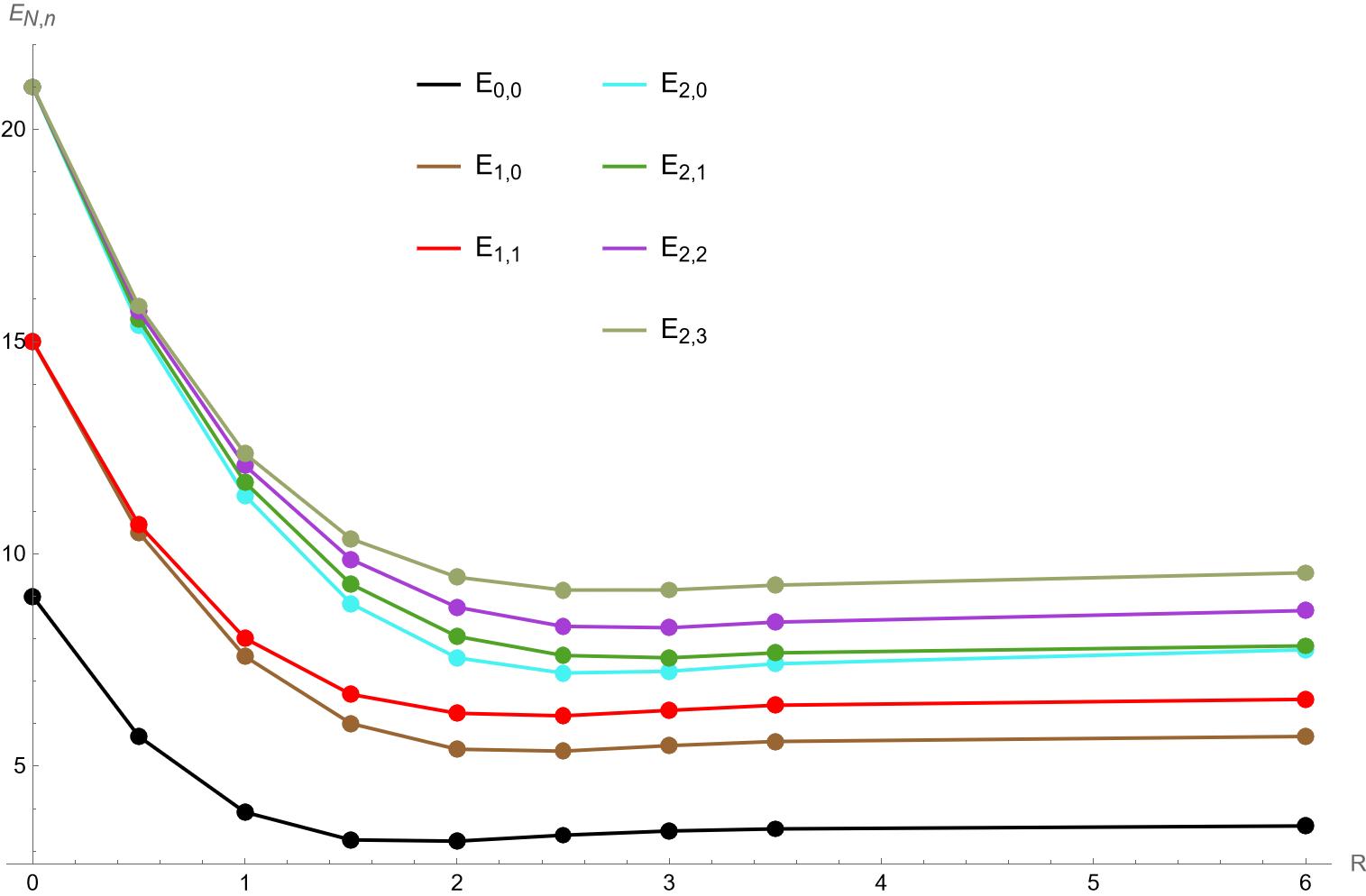}
\caption{ 
The energy $E_{N,n}$ vs $R$ of the three-body harmonic system $V_R$~(\ref{Vp}) 
for the 7 lowest $S$-states with parameters $m=1$ and $\omega=1$. It is a 
smooth function of $R$. Numerical values calculated with the Lagrange-Mesh 
method are marked by bullets, the lines are a guide to the eye (color online). Energy is presented in Hartrees.}
\label{G0}
\end{figure}

Similar calculations were carried out for the case 
$\omega =1.0$ for $R\in [0.0,6.0]$~a.u. in constant steps of $0.5$~a.u. for 
three values of $N$: $N=0, 1$ and $2$. These are depicted in 
Figure~\ref{G0}. The position of the minimum for the ground state 
$E_{0,0}$ present in Figure~\ref{G1}, appears for a rest length 
of $R_{min}\approx 1.763936$~a.u. and the energy value is 
$E_{0,0}^{min}=3.202780380949$~a.u.

Results presented in Table~\ref{t3bw05} for the system defined by the 
parameters $(m,w,R)=(1,1/2,R)$ allow us, according to the scaling 
relation~(\ref{scal}), to obtain the spectrum for the system defined by 
the parameters $(m',w',R')$ as
\begin{equation}
E\bigg[m',\,\omega',\,R'=\sqrt{\frac{1}{2\,m'\,\omega'}}\,R\bigg] = 
2\,\omega'\,E[m=1,\,\omega=1/2,\,R]\,.
\end{equation}
Likewise, for fixed $m$ and $\omega$ it can be shown that the energy tends asymptotically to a finite value at large $R$, namely $E[m,\,\omega,\,R \rightarrow \infty] \rightarrow E_{\infty}$ with $0 < E_{\infty}<E[m,\,\omega,\,R=0] $.
 
\begin{center}
\begin{table}[h]
\caption{Energy $E_{N,n}$ in a.u. of the three-body system with harmonic interactions  
for four values of $N=0,1,2,3$. Results for 11 rest lengths $R$ are
presented for $m=1$ and $\omega=0.5$.}
\scalebox{0.9}{
\begin{tabular}{l | r | rccc}
\hline\hline
$N,n$&$R=0.0$\hspace{0.6cm}\phantom{.}&$R=0.5$\hspace{0.6cm}\phantom{.}&
    $R=1.0$       & $R=1.5$       & $R=2.0$              \\
\hline
0,0&  4.500000000000&  3.248654270738& 2.398230674763& 1.893326994351& 1.658224669986\\
\hline
1,0&  7.500000000000&  5.822579071286& 4.556769365814& 3.664863226784& 3.097422523508\\
1,0&  7.500000000000&  5.822579071286& 4.556769365814& 3.664863226784& 3.097422523508\\\cline{3-6}
1,1&  7.500000000000&  5.888219366854& 4.698676482022& 3.894470971731& 3.420048994408\\
\hline
2,0& 10.500000000000&  8.425989881143& 6.767207301164& 5.497092599573& 4.585730111853\\\cline{3-6}
2,1& 10.500000000000&  8.478177133040& 6.875163036857& 5.662422650165& 4.803713591948\\
2,1& 10.500000000000&  8.478177133040& 6.875163036857& 5.662422650165& 4.803713591948\\\cline{3-6}
2,2& 10.500000000000&  8.545207941409& 7.012948079404& 5.872301106010& 5.080840901420\\
2,2& 10.500000000000&  8.545207941409& 7.012948079404& 5.872301106010& 5.080840901420\\\cline{3-6}
2,3& 10.500000000000&  8.585804667998& 7.102564535323& 6.022347722884& 5.304426211111\\
\hline
3,0& 13.500000000000& 11.117248579552& 9.151798562159& 7.580579295969& 6.377989183868\\
3,0& 13.500000000000& 11.117248579552& 9.151798562159& 7.580579295969& 6.377989183868\\\cline{3-6}
3,1& 13.500000000000& 11.166486177781& 9.256862108632& 7.749845695543& 6.621624051144\\\cline{3-6}
3,2& 13.500000000000& 11.182671178672& 9.286960974298& 7.788519430823& 6.657403147017\\\cline{3-6}
3,3& 13.500000000000& 11.201333020368& 9.327772383811& 7.855960948738& 6.756193130417\\\cline{3-6}
3,4& 13.500000000000& 11.237400167902& 9.401828686849& 7.969149439956& 6.907794387450\\
3,4& 13.500000000000& 11.237400167902& 9.401828686849& 7.969149439956& 6.907794387450\\\cline{3-6}
3,5& 13.500000000000& 11.299682142132& 9.527870507574& 8.157727630210& 7.152137692085\\
3,5& 13.500000000000& 11.299682142132& 9.527870507574& 8.157727630210& 7.152137692085\\\cline{3-6}
3,6& 13.500000000000& 11.323644849540& 9.581929109473& 8.251164113636& 7.298585516817\\ 
\hline\hline
\multicolumn{6}{r@{}}{continued \ldots}\\
\end{tabular}}
\label{t3bw05}
\end{table}
\end{center}

\begin{center}
\begin{table}[h]
\renewcommand\thetable{\ref{t3bw05}} 
\caption{Continued.}
\scalebox{0.9}{
\begin{tabular}{l | c | ccc}
\hline\hline
$N,n$&$R=2.5$      & $R=3.0$        & $R=3.5$      & $R=4.0$       \\
\hline
0,0& 1.601395264863& 1.632400069954& 1.683574737362& 1.723086021257\\
\hline
1,0& 2.792200036372& 2.676557025969& 2.674697030516& 2.718769132833\\
1,0& 2.792200036372& 2.676557025969& 2.674697030516& 2.718769132833\\\cline{2-5}
1,1& 3.190836945883& 3.103332798190& 3.091624824311& 3.131015526880\\
\hline
2,0& 4.001306153939& 3.700994915894& 3.599821980119& 3.598727725394\\\cline{2-5}
2,1& 4.252422060901& 3.946997916389& 3.809333229038& 3.770583846482\\
2,1& 4.252422060901& 3.946997916389& 3.809333229038& 3.770583846482\\\cline{2-5}
2,2& 4.579692643406& 4.292955936169& 4.152251395868& 4.120366124525\\
2,2& 4.579692643406& 4.292955936169& 4.152251395868& 4.120366124525\\\cline{2-5}
2,3& 4.884422564011& 4.669310533464& 4.577280487589& 4.566099117819\\
\hline
3,0& 5.518635696594& 4.979122899313& 4.716809070758& 4.632409629401\\
3,0& 5.518635696594& 4.979122899313& 4.716809070758& 4.632409629401\\\cline{2-5}
3,1& 5.846087784304& 5.343189633908& 5.062981319388& 4.902633799079\\\cline{2-5}
3,2& 5.857032828332& 5.382348879247& 5.105850276064& 4.953740120564\\\cline{2-5}
3,3& 5.988214691550& 5.494885196598& 5.235209305583& 5.158583646230\\\cline{2-5}
3,4& 6.173864755693& 5.701271082276& 5.401007220296& 5.222212371602\\
3,4& 6.173864755693& 5.701271082276& 5.401007220296& 5.222212371602\\\cline{2-5}
3,5& 6.458573386971& 6.007877096278& 5.730592301379& 5.579591036652\\
3,5& 6.458573386971& 6.007877096278& 5.730592301379& 5.579591036652\\\cline{2-5}
3,6& 6.675266892942& 6.308388406757& 5.857172620421& 5.654200458590\\
\hline\hline
\end{tabular}}
\label{t3bw05b}
\end{table}
\end{center}

Now, in the case of 3 particles with arbitrary masses and inspired by a phenomenological model of the
forces inside a baryon where 3 quarks
are connected by 3 glue strings and these strings meet at a
“gluon junction” (see \cite{ALEXANDROU2003667} and references therein), one can also consider a generalization of (\ref{Vp}) given by

 \begin{equation}
 \label{pot3m}
     {\tilde V}_{R}\, \equiv \, \frac{3}{2} \, \omega^{2} \,\left[ {\nu_{12}\,(r_{12}-R_{12})}^2 \, + \,  \nu_{13}\,{(r_{13}-R_{13})}^2  \, + \nu_{23}\,\,  {(r_{23}-R_{23})}^2 \right]\ ,
\end{equation}
where $\nu_{ij}$ are parameters (not all of them necessarily  positive) 
and the rest lengths $R_{ij}>0$ can take different values. Such a model 
can play an important role in molecular and atomic 3-body systems where 
the configuration of equilibrium corresponds to a triangular one (
for instance the H$_3^+$ ion~\cite{KJ:2006}). In this 
case, at $R_{12}=R_{13}=R_{23}=0$ the system is separable and exactly-solvable 
again \cite{Castro1993ExactSF} for any value of the conserved total angular 
momentum.

\clearpage

\section{Conclusions}

For the generalized 3-body harmonic system the energies of the first 14 
lowest $S-$states were computed with accuracy of 11 figures in the domain of $R\in [0.0,\,4.0]$ a.u. . At $R=0$, the problem 
becomes maximally superintegrable and exactly solvable.
The corresponding degenerate levels were analyzed in two-different 
Lie-algebraic representations. 
At $R>0$, in order to solve the Schrodinger equation for this three-body 
system~(\ref{hred}), three methods were implemented for the ground state: $i$) the perturbation 
method (PT), $ii$) the variational method (VM) and $iii$) the Lagrange-mesh 
method (LMM). The first two (PT and VM) indicated the presence of a global 
minimum in  the ground state energy for a certain value of $R=R_{\rm min}>0$, 
which was precisely confirmed by the LMM.
In all cases the energy $E(R)$ of the lowest states as a function of $R$ displays a smooth behavior 
(Figures~\ref{fENnw05} and \ref{G0}).
The degeneracy of the system when $R=0$ is partially removed for $R>0$, 
and the quantitative 
splitting of levels was presented.
Making an evident modification of the Hamiltonian (\ref{H}) the analogue 
3-body generalized harmonic system can be written for arbitrary masses 
and different spring constants~(\ref{pot3m}). Such a model can serve as 
a zero order approximation for the study of 3-body Coulomb atomic and 
molecular systems. We plan to address this subject in future studies.
It is also worth mentioning that within the LMM it is possible and straightforward
to consider states with non-zero angular momentum, $i.e.$ states with $L\ne 0$.

\begin{acknowledgements}
AMER and HOP  would like to thank A. V. Turbiner for his interest in the 
present study, helpful discussions and important remarks during its 
realization. The authors thank the KAREN cluster (ICN-UNAM, Mexico) 
where the numerical calculations were performed. 
\end{acknowledgements}

\section*{DATA AVAILABILITY}
Data sharing is not applicable to this article as no new data were created or analyzed in this
study.

\section*{Author information}
These authors contributed equally: A. M. Escobar-Ruiz and H. Olivares-Pil\'on.

\section*{Competing interests}
The authors declare no competing interests.


\clearpage

\nocite{*}

\bibliography{apssamp}

\providecommand{\noopsort}[1]{}\providecommand{\singleletter}[1]{#1}%
\begin{thebibliography}{24}
\expandafter\ifx\csname natexlab\endcsname\relax\def\natexlab#1{#1}\fi
\expandafter\ifx\csname bibnamefont\endcsname\relax
  \def\bibnamefont#1{#1}\fi
\expandafter\ifx\csname bibfnamefont\endcsname\relax
  \def\bibfnamefont#1{#1}\fi
\expandafter\ifx\csname citenamefont\endcsname\relax
  \def\citenamefont#1{#1}\fi
\expandafter\ifx\csname url\endcsname\relax
  \def\url#1{\texttt{#1}}\fi
\expandafter\ifx\csname urlprefix\endcsname\relax\def\urlprefix{URL }\fi
\providecommand{\bibinfo}[2]{#2}
\providecommand{\eprint}[2][]{\url{#2}}

\bibitem[{\citenamefont{Moshinsky and Smirnov}(1996)}]{moshinsky1996harmonic}
\bibinfo{author}{\bibfnamefont{M.}~\bibnamefont{Moshinsky}} \bibnamefont{and}
  \bibinfo{author}{\bibfnamefont{Y.}~\bibnamefont{Smirnov}},
  \emph{\bibinfo{title}{The Harmonic Oscillator in Modern Physics}},
  Contemporary concepts in physics (\bibinfo{publisher}{Harwood Academic
  Publishers}, \bibinfo{year}{1996}), ISBN \bibinfo{isbn}{9783718606214},
  \urlprefix\url{https://books.google.com.mx/books?id=RA-xtOg4z90C}.

\bibitem[{\citenamefont{Saporta~Katz and
  Efrati}(2019)}]{PhysRevLett.122.024102}
\bibinfo{author}{\bibfnamefont{O.}~\bibnamefont{Saporta~Katz}}
  \bibnamefont{and} \bibinfo{author}{\bibfnamefont{E.}~\bibnamefont{Efrati}},
  \bibinfo{journal}{Phys. Rev. Lett.} \textbf{\bibinfo{volume}{122}},
  \bibinfo{pages}{024102} (\bibinfo{year}{2019}),
  \urlprefix\url{https://link.aps.org/doi/10.1103/PhysRevLett.122.024102}.

\bibitem[{\citenamefont{Saporta~Katz and Efrati}(2020)}]{PhysRevE.101.032211}
\bibinfo{author}{\bibfnamefont{O.}~\bibnamefont{Saporta~Katz}}
  \bibnamefont{and} \bibinfo{author}{\bibfnamefont{E.}~\bibnamefont{Efrati}},
  \bibinfo{journal}{Phys. Rev. E} \textbf{\bibinfo{volume}{101}},
  \bibinfo{pages}{032211} (\bibinfo{year}{2020}),
  \urlprefix\url{https://link.aps.org/doi/10.1103/PhysRevE.101.032211}.

\bibitem[{\citenamefont{Escobar-Ruiz et~al.}(2022)\citenamefont{Escobar-Ruiz,
  Quiroz-Juarez, Del Rio-Correa, and Aquino}}]{Classicalcase}
\bibinfo{author}{\bibfnamefont{A.~M.} \bibnamefont{Escobar-Ruiz}},
  \bibinfo{author}{\bibfnamefont{M.~A.} \bibnamefont{Quiroz-Juarez}},
  \bibinfo{author}{\bibfnamefont{J.~L.} \bibnamefont{Del Rio-Correa}},
  \bibnamefont{and} \bibinfo{author}{\bibfnamefont{N.}~\bibnamefont{Aquino}},
  \emph{\bibinfo{title}{Classical harmonic three-body system: An experimental
  electronic realization}} (\bibinfo{year}{2022}),
  \urlprefix\url{https://arxiv.org/abs/2204.10501}.

\bibitem[{\citenamefont{Gutzwiller}(1971)}]{Gutzwiller1971PeriodicOA}
\bibinfo{author}{\bibfnamefont{M.~C.} \bibnamefont{Gutzwiller}},
  \bibinfo{journal}{Journal of Mathematical Physics}
  \textbf{\bibinfo{volume}{12}}, \bibinfo{pages}{343} (\bibinfo{year}{1971}).

\bibitem[{\citenamefont{Richard}(1992)}]{JMR:1992}
\bibinfo{author}{\bibfnamefont{J.-M.} \bibnamefont{Richard}},
  \bibinfo{journal}{Phys. Rep.} \textbf{\bibinfo{volume}{212}},
  \bibinfo{pages}{1} (\bibinfo{year}{1992}).

\bibitem[{\citenamefont{Turbiner et~al.}(2017)\citenamefont{Turbiner, Miller,
  and Escobar-Ruiz}}]{Turbiner_2017}
\bibinfo{author}{\bibfnamefont{A.~V.} \bibnamefont{Turbiner}},
  \bibinfo{author}{\bibfnamefont{W.}~\bibnamefont{Miller}}, \bibnamefont{and}
  \bibinfo{author}{\bibfnamefont{A.~M.} \bibnamefont{Escobar-Ruiz}},
  \bibinfo{journal}{Journal of Physics A: Mathematical and Theoretical}
  \textbf{\bibinfo{volume}{50}}, \bibinfo{pages}{215201}
  (\bibinfo{year}{2017}),
  \urlprefix\url{https://doi.org/10.1088/1751-8121/aa6cc2}.

\bibitem[{\citenamefont{Turbiner et~al.}(2018)\citenamefont{Turbiner, Miller,
  and Escobar-Ruiz}}]{doi:10.1063/1.4994397}
\bibinfo{author}{\bibfnamefont{A.~V.} \bibnamefont{Turbiner}},
  \bibinfo{author}{\bibfnamefont{W.}~\bibnamefont{Miller}}, \bibnamefont{and}
  \bibinfo{author}{\bibfnamefont{M.~A.} \bibnamefont{Escobar-Ruiz}},
  \bibinfo{journal}{Journal of Mathematical Physics}
  \textbf{\bibinfo{volume}{59}}, \bibinfo{pages}{022108}
  (\bibinfo{year}{2018}), \eprint{https://doi.org/10.1063/1.4994397},
  \urlprefix\url{https://doi.org/10.1063/1.4994397}.

\bibitem[{\citenamefont{Gu et~al.}(2002)\citenamefont{Gu, Duan, and Ma}}]{Gu}
\bibinfo{author}{\bibfnamefont{X.-Y.} \bibnamefont{Gu}},
  \bibinfo{author}{\bibfnamefont{B.}~\bibnamefont{Duan}}, \bibnamefont{and}
  \bibinfo{author}{\bibfnamefont{Z.-Q.} \bibnamefont{Ma}},
  \bibinfo{journal}{Journal of Mathematical Physics}
  \textbf{\bibinfo{volume}{43}}, \bibinfo{pages}{2895} (\bibinfo{year}{2002}),
  \eprint{https://doi.org/10.1063/1.1476393},
  \urlprefix\url{https://doi.org/10.1063/1.1476393}.

\bibitem[{\citenamefont{Fernández}(2008)}]{FF}
\bibinfo{author}{\bibfnamefont{F.~M.} \bibnamefont{Fernández}}
  (\bibinfo{year}{2008}), \urlprefix\url{https://arxiv.org/abs/0810.2210}.

\bibitem[{\citenamefont{de~Castro and Sugaya}(1993)}]{Castro1993ExactSF}
\bibinfo{author}{\bibfnamefont{A.~S.} \bibnamefont{de~Castro}}
  \bibnamefont{and} \bibinfo{author}{\bibfnamefont{M.}~\bibnamefont{Sugaya}},
  \bibinfo{journal}{European Journal of Physics} \textbf{\bibinfo{volume}{14}},
  \bibinfo{pages}{259} (\bibinfo{year}{1993}).

\bibitem[{\citenamefont{Turbiner
  et~al.}(2020{\natexlab{a}})\citenamefont{Turbiner, Miller, and
  Escobar-Ruiz}}]{Turbiner_2020}
\bibinfo{author}{\bibfnamefont{A.~V.} \bibnamefont{Turbiner}},
  \bibinfo{author}{\bibfnamefont{W.}~\bibnamefont{Miller}}, \bibnamefont{and}
  \bibinfo{author}{\bibfnamefont{M.~A.} \bibnamefont{Escobar-Ruiz}},
  \bibinfo{journal}{Journal of Physics A: Mathematical and Theoretical}
  \textbf{\bibinfo{volume}{53}}, \bibinfo{pages}{055302}
  (\bibinfo{year}{2020}{\natexlab{a}}),
  \urlprefix\url{https://doi.org/10.1088/1751-8121/ab5f39}.

\bibitem[{\citenamefont{{Znojil}}(2003)}]{Znojil}
\bibinfo{author}{\bibfnamefont{M.}~\bibnamefont{{Znojil}}},
  \bibinfo{journal}{Nuclear Physics B} \textbf{\bibinfo{volume}{662}},
  \bibinfo{pages}{554} (\bibinfo{year}{2003}), \eprint{hep-th/0209262}.

\bibitem[{\citenamefont{Delves}(1960)}]{LMD:1960}
\bibinfo{author}{\bibfnamefont{L.~M.} \bibnamefont{Delves}},
  \bibinfo{journal}{Nucl. Phys. E} \textbf{\bibinfo{volume}{20}},
  \bibinfo{pages}{275} (\bibinfo{year}{1960}),
  \urlprefix\url{https://www.sciencedirect.com/science/article/abs/pii/0029558260901747}.

\bibitem[{\citenamefont{Turbiner
  et~al.}(2020{\natexlab{b}})\citenamefont{Turbiner, Miller, and
  Escobar-Ruiz}}]{From2dimensional}
\bibinfo{author}{\bibfnamefont{A.~V.} \bibnamefont{Turbiner}},
  \bibinfo{author}{\bibfnamefont{W.}~\bibnamefont{Miller}}, \bibnamefont{and}
  \bibinfo{author}{\bibfnamefont{M.~A.} \bibnamefont{Escobar-Ruiz}},
  \bibinfo{journal}{Journal of Physics A: Mathematical and Theoretical}
  \textbf{\bibinfo{volume}{54}}, \bibinfo{pages}{015204}
  (\bibinfo{year}{2020}{\natexlab{b}}),
  \urlprefix\url{https://doi.org/10.1088/1751-8121/abcb43}.

\bibitem[{\citenamefont{Turbiner}(1984)}]{NLP}
\bibinfo{author}{\bibfnamefont{A.~V.} \bibnamefont{Turbiner}},
  \bibinfo{journal}{Soviet Physics Uspekhi} \textbf{\bibinfo{volume}{27}},
  \bibinfo{pages}{668} (\bibinfo{year}{1984}).

\bibitem[{\citenamefont{Baye and Heenen}(1986)}]{BH:1986}
\bibinfo{author}{\bibfnamefont{D.}~\bibnamefont{Baye}} \bibnamefont{and}
  \bibinfo{author}{\bibfnamefont{P.~H.} \bibnamefont{Heenen}},
  \bibinfo{journal}{J. Phys. A: Math. Gen.} \textbf{\bibinfo{volume}{19}},
  \bibinfo{pages}{2041} (\bibinfo{year}{1986}).

\bibitem[{\citenamefont{Baye}(2015)}]{DB:2015}
\bibinfo{author}{\bibfnamefont{D.}~\bibnamefont{Baye}}, \bibinfo{journal}{Phys.
  Rep.} \textbf{\bibinfo{volume}{32}}, \bibinfo{pages}{1–107}
  (\bibinfo{year}{2015}).

\bibitem[{\citenamefont{Pekeris}(1958)}]{Pekeris1958}
\bibinfo{author}{\bibfnamefont{C.~L.} \bibnamefont{Pekeris}},
  \bibinfo{journal}{Physical Review} \textbf{\bibinfo{volume}{112}},
  \bibinfo{pages}{1649} (\bibinfo{year}{1958}).

\bibitem[{\citenamefont{Baye and Dohet-Eraly}(2015)}]{BJ:2015}
\bibinfo{author}{\bibfnamefont{D.}~\bibnamefont{Baye}} \bibnamefont{and}
  \bibinfo{author}{\bibfnamefont{J.}~\bibnamefont{Dohet-Eraly}},
  \bibinfo{journal}{Phys. Chem. Chem. Phys.} \textbf{\bibinfo{volume}{17}},
  \bibinfo{pages}{31417} (\bibinfo{year}{2015}),
  \urlprefix\url{https://pubs.rsc.org/en/content/articlelanding/2015/cp/c5cp00110b}.

\bibitem[{\citenamefont{Hesse and Baye}(2001)}]{HB:2001}
\bibinfo{author}{\bibfnamefont{M.}~\bibnamefont{Hesse}} \bibnamefont{and}
  \bibinfo{author}{\bibfnamefont{D.}~\bibnamefont{Baye}}, \bibinfo{journal}{J.
  Phys. B: At. Mol. Opt. Phys.} \textbf{\bibinfo{volume}{34}},
  \bibinfo{pages}{1425–1442} (\bibinfo{year}{2001}),
  \urlprefix\url{https://iopscience.iop.org/article/10.1088/0953-4075/34/8/308}.

\bibitem[{\citenamefont{Hesse and Baye}(2003)}]{HB:2003}
\bibinfo{author}{\bibfnamefont{M.}~\bibnamefont{Hesse}} \bibnamefont{and}
  \bibinfo{author}{\bibfnamefont{D.}~\bibnamefont{Baye}}, \bibinfo{journal}{J.
  Phys. B: At. Mol. Opt. Phys.} \textbf{\bibinfo{volume}{36}},
  \bibinfo{pages}{139–154} (\bibinfo{year}{2003}),
  \urlprefix\url{https://iopscience.iop.org/article/10.1088/0953-4075/36/1/311}.

\bibitem[{\citenamefont{Alexandrou et~al.}(2003)\citenamefont{Alexandrou, {de
  Forcrand}, and Jahn}}]{ALEXANDROU2003667}
\bibinfo{author}{\bibfnamefont{C.}~\bibnamefont{Alexandrou}},
  \bibinfo{author}{\bibfnamefont{P.}~\bibnamefont{{de Forcrand}}},
  \bibnamefont{and} \bibinfo{author}{\bibfnamefont{O.}~\bibnamefont{Jahn}},
  \bibinfo{journal}{Nuclear Physics B - Proceedings Supplements}
  \textbf{\bibinfo{volume}{119}}, \bibinfo{pages}{667} (\bibinfo{year}{2003}),
  ISSN \bibinfo{issn}{0920-5632}, \bibinfo{note}{proceedings of the XXth
  International Symposium on Lattice Field Theory},
  \urlprefix\url{https://www.sciencedirect.com/science/article/pii/S0920563203016591}.

\bibitem[{\citenamefont{Kutzelnigg and Jaquet}(2006)}]{KJ:2006}
\bibinfo{author}{\bibfnamefont{W.}~\bibnamefont{Kutzelnigg}} \bibnamefont{and}
  \bibinfo{author}{\bibfnamefont{R.}~\bibnamefont{Jaquet}},
  \bibinfo{journal}{Phil. Trans. R. Soc. A} \textbf{\bibinfo{volume}{364}},
  \bibinfo{pages}{2855–2876} (\bibinfo{year}{2006}),
  \urlprefix\url{https://royalsocietypublishing.org/doi/full/10.1098/rsta.2006.1871}.

\end{thebibliography}

\end{document}